\pdfoutput=1
\documentclass[preprint,svgnames,dvipsnames,authoryear]{elsarticle}

\usepackage[OT1]{fontenc}
\usepackage{amsmath,amsfonts, amssymb,amsthm}
\usepackage{natbib}
\usepackage[colorlinks,urlcolor=blue]{hyperref}
\usepackage[left=3cm,right=3cm,top=3cm,bottom=3cm]{geometry}
\usepackage{algorithm}
\usepackage{algorithmicx}
\usepackage{algpseudocode}
\usepackage{bm}
\usepackage{graphicx}
\usepackage{dsfont}
\usepackage{tabularx}
\usepackage{stmaryrd}
\usepackage[svgnames,dvipsnames]{xcolor}
\usepackage{listings}
\usepackage{amssymb}
\usepackage{booktabs}
\usepackage{enumerate}

\newcommand{\pset}[1]{\mathcal{P}(#1)}

\makeatletter
\def\ps@pprintTitle{%
\let\@oddhead\@empty
\let\@evenhead\@empty
\def\@oddfoot{}%
\let\@evenfoot\@oddfoot}
\makeatother

\title{Shapley effects and proportional marginal effects for global sensitivity analysis: application to computed tomography scan organ dose estimation}

\author[a]{Anaïs Foucault}
\author[b,c,d]{Marouane Il Idrissi}
\author[b,c,d]{Bertrand Iooss}
\author[a,e]{Sophie Ancelet}

\address[a]{Institut de Radioprotection et de Sureté Nucléaire (IRSN), PSE-SANTE/SESANE/LEPID, BP 17, 92262 Fontenay-aux-Roses, France}
\address[b]{EDF Lab Chatou, 6 Quai Watier, 78401 Chatou, France}
\address[c]{SINCLAIR AI Lab., Saclay, France}
\address[d]{Institut de Mathématiques de Toulouse, 31062 Toulouse, France}
\address[e]{Corresponding Author -  Email: \tt{sophie.ancelet@irsn.fr}}

\begin{document}
\begin{frontmatter}
\begin{abstract} 
Concerns have been raised about possible cancer risks after exposure to computed tomography (CT) scans in childhood. The health effects of ionizing radiation are then estimated from the absorbed dose to the organs of interest which is calculated, for each CT scan, from dosimetric numerical models, like the one proposed in the NCICT software. Given that a dosimetric model depends on input parameters which are most often uncertain, the calculation of absorbed doses is inherently uncertain. A current methodological challenge in radiation epidemiology is thus to be able to account for dose uncertainty in risk estimation. A preliminary important step can be to identify the most influential input parameters implied in dose estimation, before modelling and accounting for their related uncertainty in radiation-induced health risks estimates. In this work, a variance-based global sensitivity analysis was performed to rank by influence the uncertain input parameters of the NCICT software implied in brain and red bone marrow doses estimation, for four classes of CT examinations. Two recent sensitivity indices, especially adapted to the case of dependent input parameters, were estimated, namely: the Shapley effects and the Proportional Marginal Effects (PME). This provides a first comparison of the respective behavior and usefulness of these two indices on a real medical application case. The conclusion is that Shapley effects and PME are intrinsically different, but complementary. Interestingly, we also observed that the proportional redistribution property of the PME allowed for a clearer importance hierarchy between the input parameters.
\end{abstract}

\end{frontmatter}

\section{Introduction}
\label{intro}

In radiation epidemiology, one is commonly interested in the potential association between health risks and cumulative exposure to ionizing radiation (IR) \citep{BEIRVII,Unscear:2012}. Particularly, concerns have been raised about possible cancer risks after exposure to computed tomography (CT) scans in childhood \citep{Unscear:2013}. In this specific context, individual exposure conditions to IR cannot be controlled experimentally: they are thus inherently uncertain. Additionally, the health effects of IR are associated with the absorbed dose to the organs of interest, rather than with radiation exposure. The absorbed dose does not only depend on the exposure to radioactive material but also on individual exposure conditions parameters. An absorbed dose is not directly measurable. It is calculated, for each CT scan, from numerical  models, like the one proposed in the NCICT software \citep{Lee:2015}. Given that a dosimetric model depends on input parameters which are most often uncertain (e.g. individual exposure conditions to IR), the calculation of absorbed doses is uncertain when estimating radiation-induced health risks after CT scans. However, when it is not or only poorly accounted for, dose uncertainty can lead to biased risk estimates, a loss in statistical power and to a distortion of the dose-response relationship \citep{Thomas:1993,Kim:2006,Physick:2007}. Therefore, a current methodological challenge in radiation epidemiology is to be able to account for dose uncertainty in risk estimation \citep{NCRP:171}. In this context, a preliminary important step can be to identify the most influential input parameters implied in dose estimation, before modelling and accounting for their related uncertainty in radiation-induced health risks estimates.

For that purpose, global sensitivity analysis (GSA) provides tools to quantify the influence of uncertain inputs of a numerical model (often assumed to be black-box) \citep{Saltelli:2008,borpli16,borrab23}. It is one of the key steps in uncertainty quantification of numerical models, especially in the context of scientific and industrial practices \citep{de_rocquigny:2008}, and in particular in operation research and managerial problems \citep{bor17}. \citet{daVeiga:2021livre} distinguish four major settings addressed by GSA: (i.) model exploration, i.e., investigating the input-output relationship; (ii.) factor fixing, i.e., identifying non-influential inputs; (iii.) factor prioritization, i.e., quantifying the most important inputs using quantitative importance measures; (iv.) robustness analysis, i.e., quantifying the sensitivity of the quantity of interest with respect to probabilistic model uncertainty of the input distributions. In the present work, the third setting is of interest (i.e., factor prioritization), given that our aim is to identify and rank the most influential input parameters of the NCICT software for organ dose estimation.

Among the large panel of GSA indices, the well-known Sobol’ indices \citep{Sobol:1993} are derived from the functional analysis of variance (FANOVA) decomposition \citep{Efron:1981}. This decomposition relies on the assumption of mutual independence between the inputs and attributes percentages of the model output variance to every possible subset of inputs. These indices sum to one, are non-negative, and can be interpreted as individual/interaction effects on the output. However, the mutual independence assumption can, in some cases, be unrealistic. In many applications, inputs have an inherent statistical dependence structure, either imposed in their probabilistic modeling or induced by physical constraints upon the input or the output space. This is particularly the case for some of the input parameters of the NCICT software. When the inputs are dependent, the Sobol’ indices lose their intrinsic interpretation as individual/interaction effects, (see, e.g., \citet{daVeiga:2021livre} for an overview on the different works dealing with this subject).

Among the different approaches proposed to circumvent this issue, the Shapley effects have been shown to provide synthetic and relevant information about the part of the model output variance that is due to each input in many practical cases (see, e.g., \citet{IoossPrieur:2019,radsur19}), and especially in epidemiological modeling  (see, e.g., \citet{daVeiga:2021livre,Idrissi:2021,davilapena:2022}).
Based on the Shapley value \citep{shapley:1953} and inspired from cooperative game theory \citep{Osborne:1994}, the Shapley effects have been introduced by \citet{Owen:2014} in GSA. The underlying idea is to consider the inputs as players of a cooperative game, to which the variance of the model output must be redistributed. 
This method attributes non-negative shares of variance to every input, and they always sum to one, despite the potential dependence structure. Hence, their interpretation is always guaranteed. These indices and their estimation methods have been largely studied in the GSA literature \citep{OwenPrieur:2017, beneli19, plirab21,rabbor19,Song:2016,Broto:2020}.

However, for certain purposes, the Shapley effects can be criticized for one of its side effects: they may attribute a non-zero share of the output variance to an exogenous input (i.e., an input that does not appear in the deterministic model) that is correlated with an influential one \citep{IoossPrieur:2019}. This phenomenon, which is synonym to the violation of the so-called ``exclusion property'' of an importance measure \citep{Gromping:2007,ioocha22,cloioo23}, is sometimes called the Shapley's joke \citep{ioocha22} or the correlation distorsion \citep{verdinelli:2023}.
In order to prevent this phenomenon, \citet{Herin:2022} have recently defined sensitivity indices using another allocation system, namely the proportional values \citep{Ortmann:2000, Feldman:2005}. In the context of variance-based GSA, it leads to the definition of novel indices called the proportional marginal effects (PME). Like the Shapley effects, they always remain interpretable even with a potential dependence structure between input parameters, but they also allow to detect exogenous inputs by granting them zero allocation, despite their eventual correlation with non-exogenous inputs.

Although Shapley effects and PME are promising variance-based GSA indices for forthcoming applications in many research areas, they have not yet been widely used in health studies and have not been used together to highlight their potential complementary. Thus, as pointed out by \citet{Lu:2023}, in the particular context of pandemic models like the susceptible-infectious-recovered (SIR) models, sensitivity analysis and uncertainty quantification are rarely performed. Moreover, the most widely employed methods are one-at-a-time sensitivity analysis omitting the potential interactions between input parameters. As an example, \citet{Wang:2018} focused on Sobol' indices to evaluate the relative importance of multiple input parameters when developing a mathematical model to replace invasive and time-consuming biological measures in the management of patients with chronic liver diseases. Likewise, \citet{Rapadamnaba:2021} evaluated a cardiovascular model of patient-specific arterial network. \citet{Roder:2021} highlighted the usefulness of the Shapley values for assessing the relative importance of the input features of a cohort of patients in the context of a machine learning (ML) algorithm trained to predict the result of a molecular diagnostic test at the sample or patient level. In the same way, \citet{smith2021} used Shapley values to interpret ML models and identify mortality factors for COVID-19. 

In this work, a variance-based GSA is performed with both Shapley effects and PME and, for the first time, in radiation dosimetry, to rank by influence the input parameters of the NCICT software classically used to estimate the organ doses arising from CT scan exposure. Their respective behavior and usefulness are highlighted through this medical application.

The structure of the paper is as follows. Section \ref{study} describes the case-study. Section \ref{gsa} briefly presents Shapley effects and PME for variance-based GSA and their practical implementation. These two indices are first compared in Section \ref{appli_covid} in the context of a use-case example based on a COVID-19 SIR model. In Section \ref{appli_CT}, they are applied to the CT organ dose estimation model. Finally, our results are discussed in Section \ref{discu}.

\section{Case study: CT scan organ dose estimation from the NCICT software}
\label{study}

To estimate the organ doses associated to a CT scan, the NCICT computer program \citep{Lee:2015} is classically used \citep{Lee:2019,ThierryChef:2021,Foucault:2022}. NCICT uses a library of phantoms, developed by the University of Florida/National Cancer Institute and the International Commission on Radiological Protection (ICRP). These phantoms represent realistic human anatomy \citep{Menzel:2009, Bolch:2020}. In this work, we focus on the version 1.0 of the NCICT software for which the organ dose estimation following CT scan is possible for 12 phantoms (females and males, six ages: newborn, 1, 5, 10, 15 years old and adult).
The (deterministic) numerical model implemented in NCICT to estimate organ doses is defined by the following input parameters: age and gender of the patient, scanned body region (identified by the scan start and stop landmarks measured in cm from the top of the head), CT machine model and technical parameters used for each image acquisition including tube current-time product (mAs), tube potential (kVp), pitch and bowtie filter (16 cm - head or 32 cm - body). 
Head filter is systematically assigned to head scans whereas body filter is considered for the other types of examinations. Table \ref{inputs} describes the eight input parameters of the NCICT dosimetric model that we consider in this work; the GSA problem is then of dimension $d=8$.

\begin{table}[!ht]
\centering
\begin{tabular}{c c c c} 
\hline
Input & Description & Type of variable & Range of values \\ 
\hline
Age & Age of the patient & Discrete & $\left\{0, \ldots, 18\right\}$\\
Gender & Gender of the patient & Categorical & $\left\{F,M\right\}$\\
Start & Start of the scanned body area (in cm) & Discrete & $\left\{1, \ldots, 165\right\}$\\
End & End of the scanned body area (in cm) & Discrete & $\left\{1, \ldots, 165\right\}$\\
mAs & Tube current-time product & Discrete & $\left\{4, \ldots, 505\right\}$\\
kVp & Tube potential & Discrete & $\left\{80,100,120,140\right\}$\\
pitch & Pitch & Continuous & $\left[0.2, \ldots, 1.75\right]$\\
Model & CT model & Categorical & $\left\{1, \ldots, 12\right\}$\\
\hline
\end{tabular}
\caption{Description of the input parameters of the NCICT (version 1.0) dosimetric model used for CT scan dose estimation.}
\label{inputs}
\end{table}

In NCICT version 1.0, the absorbed dose (in mGy) related to an $organ$ of interest and following a CT scan received by a patient with a given $age$ and $gender$ is defined by the following equation:
\begin{equation}
\begin{array}{l}
    D(organ, age, gender, model, kVp, filter, start, end, mAs, pitch) \\
    \displaystyle = \sum_{z=start}^{end} CTDI_{vol}(model,kVp,filter,mAs,pitch) \times DC(organ,age,gender,kVp,filter,z) .
    \label{NCICTequation}
    \end{array}
\end{equation}
where:
\begin{itemize}
    \item $z$ is a 1 cm axial slice between $start$ and $end$ from the top of the head.
        \item $CTDI_{vol}(model,kVp,filter,mAs,pitch)$ is the CT Dose Index per unit volumetric, obtained for the particular scanner model studied and the technical parameters used for image acquisition. It is derived from the following equation:
\begin{equation}
    CTDI_{vol}(model,kVp,filter,mAs,pitch)=\frac{nCTDI_w(model,kVp,filter)}{pitch} \times \frac{mAs}{100},
    \label{CTDIvol}
\end{equation}
where $nCTDI_w(model,kVp,filter)$ is the $CTDI_w$ normalised to $100$ mAs and selected from the CTDI library \citep{Lee:2014}.

    \item $DC(organ,age,gender,kVp,filter,z)$ is an organ dose coefficient for $z$, depending on patient characteristics and technical acquisition parameters. 
    NCICT includes Monte Carlo simulations (MCNPX2.7) of the x-ray in a reference CT scanner (Siemens Sensation 16) to estimate organ doses coefficients ($DC$). In this work, $DC$ were assumed to be known for each possible configuration of input parameters.

\end{itemize}

Note that, in this work, we only used a  "restricted" version of the dosimetric model implemented in the NCICT software (version 1.0) (called "restricted" NCICT model hereafter). Indeed, the black-box Monte Carlo simulations model included in NCICT to calculate the organ doses coefficients $DC$ was ignored since only a library of $DC$ values was available.
To perform a variance-based GSA from the NCICT dosimetric model, we used an input-output sample of PACS (Picture Archiving and Communication System) data, available for some French university hospitals and some pediatric patients. PACS records and archives all images and machine settings of CT examinations. For each CT image recorded in PACS, information on the patient (gender, date of birth) and on the examination (scan date, scanned body region, CT machine model and machine settings) were collected.

\section{Variance-based GSA with dependent inputs: Shapley effects and PME}
\label{gsa}

This section presents the formal framework of variance-based GSA, with a special focus on the Shapley effects and the PME. Let $\mathbf{X}=(X_1, \dots, X_d)$ be a random vector, where each $X_i, i \in D:= \{1, \dots, d\}$ represents an uncertain input of a real-valued deterministic model, denoted $f$ hereafter. Denote $\mathcal{P}(D)$ the \emph{power-set} of $D$, i.e., the set of subsets of $D$. For any $A \in \pset{D}$, let $X_A:= (X_i)_{i \in A}$ denotes the subset of the inputs defined in $\mathbf{X}$ and 
whose indices are in $A$. Moreover, the complementary of a subset $A \in \pset{D}$ in $D$ is denoted $-A := D \setminus A$. $|~.~|$ denotes the number of elements in a subset. The real-valued output random variable of $f$ is denoted $Y = f(X)$. Let $\mathbb{E}(\cdot)$ and $\mathbb{V}(\cdot)$ denote the expectation and variance operators, respectively. In the following, when a function is referred to as being non-negative (resp. positive), it entails that it is valued in $[0, \infty)$ (resp. $(0, \infty)$).

\subsection{Hoeffding's functional ANOVA and the Sobol' indices}

\begin{figure}[ht!]
    \centering
    \includegraphics[width=0.4\textwidth]{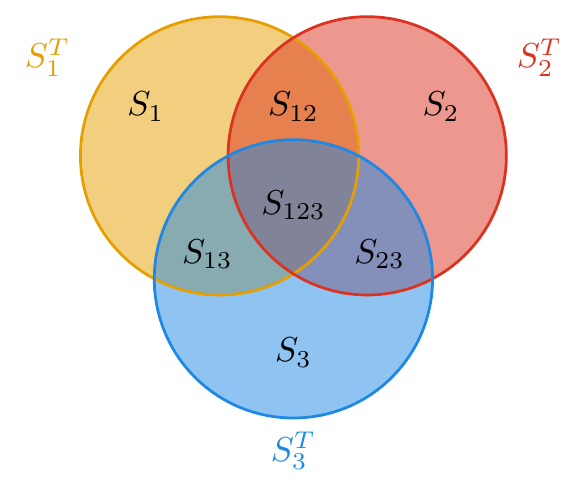}
    \caption{Illustration of the classical and total Sobol' indices of three mutually independent inputs.}
    \label{fig:sch_sobol}
\end{figure}

If the random vector of inputs $\mathbf{X}$ is assumed to be composed of mutually independent random variables, and if $\mathbb{E}(Y^2)$ is assumed to be finite, one can write \citep{daVeiga:2021livre}:
$$\mathbb{V}(Y) = \sum_{A \in \pset{D}} V_A,$$
where, $\forall A \in \pset{D}$,
$$V_A = \sum_{B \in \pset{A}} (-1)^{|A|-|B|} \mathbb{V}\left( \mathbb{E}\left( Y \mid X_B\right)\right).$$
This result is due to a unique additive orthogonal decomposition of $f$ in the (Hilbert) space of square-integrable functions \citep{Hoeffding:1948,Efron:1981}. The (classical) Sobol' indices have then been defined, $\forall A \in \pset{D}$ as \citep{Sobol:1993}:
\begin{equation}
    S_A := \frac{V_A}{\mathbb{V}(Y)}.
    \label{eq:sobol}
\end{equation}
Related to this decomposition, other indices have also been proposed: the \emph{total} Sobol' indices. They are defined, $\forall A \in \pset{D}$, as:
\begin{equation}
    S^T_A = \frac{\mathbb{E}[\mathbb{V}(Y|X_{-A})]}{\mathbb{V}(Y)} = \sum_{B \in \pset{A}} S_B.
    \label{eq:totalsobol}
\end{equation}
Whenever the inputs are mutually independent, the Sobol' indices of each input quantify their individual influence, and for each subset of several inputs, their related Sobol' index quantifies their interaction due to the model $f$. The total Sobol' indices of a subset of inputs can thus be interpreted as the sum of individual/interaction influences (i.e., the Sobol' indices) of every subset of this subset. Figure~\ref{fig:sch_sobol} illustrates the variance decomposition in the case of three mutually independent inputs. The overall area of this Venn diagram represents $\mathbb{V}(Y)$, each colored areas represent the Sobol' indices, and each colored circle represents the total Sobol' indices of each input.

It is important to note that, even if the Sobol' indices in Eq.~(\ref{eq:sobol}) and total Sobol' indices in Eq.~(\ref{eq:totalsobol}) remain well-defined (i.e., finite) as long as $Y$ has a finite variance, their interpretation changes whenever the input mutual independence is no longer assumed. As stated in \citet{daVeiga:2021livre}, whenever the inputs are correlated, the Sobol' indices as defined in Eq.~(\ref{eq:sobol}) can become negative, and their interpretation as percentages of variance (and hence as influence) can be challenged.

\subsection{Allocations and random order models}

Cooperative (or coalitional) game theory is a field coming from economics, which primarily deals with the redistribution of wealth \citep{Osborne:1994}. Formally, the game is composed of a set $D=\{1,\dots,d\}$ whose elements represent players, and a user-chosen value function $v : \pset{D} \rightarrow \mathbb{R}$, whose purpose is to quantify the value of a coalition (i.e., subset) of players $A \in \pset{D}$. The total quantity to be redistributed 
among all players is the value of the \emph{grand coalition}: $v(D)$.

To that extent, \emph{allocations} (or solution concepts) have been proposed, resulting in a redistribution of $v(D)$, where each player $i \in D$ receives a share $\phi_i$. The main way different allocations can be compared is through their properties (i.e., axioms). One famous example of allocation is the Shapley values \citep{shapley:1953}. They can be defined, $\forall i \in D$, as:
\begin{align}
    \textrm{Shap}_i &= \sum_{A \in \pset{D} ; i \in A} \frac{\sum_{B \in \pset{A}} (-1)^{|A|-|B|} v(B)}{|A|} \label{eq:shapHarsanyi}\\
    &=\frac{1}{d!}\sum_{\pi \in S_D} \left[v\left(C_{\pi(i)}(\pi)\right)-v\left(C_{\pi(i)-1}(\pi)\right)\right] \label{eq:shapWeber}
\end{align}
where $S_D$ is the set of all permutations of $D$, and for a particular permutation $\pi=\left(\pi_1,\ldots,\pi_d\right) \in S_D$, $\pi(i)=\pi_i^{-1}$ denotes the position of the player $i$ in $\pi$ (i.e., $\pi_{\pi(i)}=i$), and finally $C_i(\pi)$ denotes the set of the $i$-th first inputs in the ordering $\pi$ such as $C_i(\pi)=\left\{\pi_j : j \leq i\right\}$ with the convention that $C_0(\pi)={\varnothing}$. Eq.~(\ref{eq:shapHarsanyi}) is due to \citet{Harsanyi1963}, whereas Eq.~(\ref{eq:shapWeber}) is an equivalent expression of the Shapley values as a random order model allocation \citep{Weber:1988}.

Random order model allocations are a set of allocations that can be written generally as:
\begin{equation}
    \label{order_alloc}
    \phi_i=\sum_{\pi \in S_D} p(\pi)\left[v\left(C_{\pi(i)}(\pi)\right)-v\left(C_{\pi(i)-1}(\pi)\right)\right]
\end{equation}
where $p$ is a probability mass function over the orderings of $D$. 
In this framework, the Shapley values can be understood as a (discrete) uniform choice for $p$. Two remarkable properties arise from random order model allocations (whatever the choice of $p$ is) \citep{Weber:1988}:
\begin{itemize}
    \item They are \emph{efficient:} $\sum_{i=1}^d \phi_i = v(D)$;
    \item If $v$ is chosen to be monotonic (i.e., if $B \subseteq A \in \pset{D}$ then $v(B) \leq v(A)$), then they are \emph{non-negative}: $\phi_i \geq 0$.
\end{itemize}

Different choices of probability mass function $p$ lead to different allocations. In particular, the \emph{proportional values} \citep{Ortmann:2000} result from the choice:
$$p(\pi)=\frac{L(\pi)}{\sum_{\sigma \in S_D}L(\sigma)}, \quad \text{where}\quad L(\pi)=\left(\prod_{j \in D} v(C_j(\pi))\right)^{-1}.$$

Cooperative games can be endowed with a dual \citep{Feldman:2005, Feldman:2007}. Instead of focusing on the value brought in by a coalition, the dual formulation instead defines a cooperative game based on their \emph{worth}. It implies defining the function, $\forall A \in \pset{D}$, as:
$$w(A) = v(D) - v(-A),$$
and computing the allocations by substituting the value function $v$ by $w$. In the context of random order models, as pointed out in \citet{Gromping:2007} and \citet{Herin:2022}, this different point of view can be interpreted as quantifying the sequential costs of removing players from coalitions (for the dual game) instead of quantifying the sequential gains of adding them (for the initial game). It is interesting to note that the Shapley values of a cooperative game are equivalent to the Shapley values of its dual \citep{Funaki1996}. However, this behavior is not respected for the proportional values \citep{Feldman:2007}.

\subsection{Variance-based GSA indices inspired by allocations}

By analogy between players in a cooperative game and inputs of a deterministic numerical model, \citet{Owen:2014} proposed the Shapley effects as measures of influence whenever the inputs are correlated. These indices are none other than the Shapley values of the cooperative game defined with the value function:
$$v(A) = \frac{\mathbb{V}(\mathbb{E}(Y \mid X_A))}{\mathbb{V}(Y)}.$$
The Shapley effects can thus be written, coming from Eq.~(\ref{eq:shapHarsanyi}), $\forall i \in D$, as:
\begin{align}
    \textrm{Sh}_i &= \sum_{A \in \pset{D} ; i \in A} \frac{\sum_{B \in \pset{A}}(-1)^{|A|-|B|}\mathbb{V}(\mathbb{E}(Y \mid X_B))}{\mathbb{V}(Y) |A|} \notag \\
    &= \sum_{A \in \pset{D} ; i \in A} \frac{S_A}{|A|}, \label{eq:shHarsanyi}
\end{align}
where $S_A$ are the Sobol' indices as defined in Eq~(\ref{eq:sobol}), but potentially computed with dependent inputs. 
These indices have been extensively studied both empirically and theoretically in the literature \citep{OwenPrieur:2017, beneli19, plirab21,Song:2016,Broto:2020,IoossPrieur:2019, Idrissi:2021}. One can notice that the dual of this game leads, $\forall A \in \pset{D}$, to:
$$w(A) = S^T_A,$$
and, since the Shapley values of a game are equal to the Shapley values of its dual, it leads to the same indices. This fact has also been noticed by \citet{Song:2016} in the field of sensitivity analysis. Hence, another equivalent rewriting of the Shapley effects, for any $i \in D$, is:
$$\textrm{Sh}_i = \sum_{A \in \pset{D} ; i \in A} \frac{\sum_{B \in \pset{A}}(-1)^{|A|-|B|}S^T_B}{|A|} .$$

The Shapley effects have two remarkable properties, making them particularly attractive for influence quantification with dependent inputs:
\begin{itemize}
    \item They sum to $1$;
    \item They are always non-negative.
\end{itemize}
Hence, they can be interpreted as percentages of variance attributed to each input.

However, the Shapley effects suffer from one main drawback: if exogenous inputs (i.e., not in the deterministic model) are sufficiently correlated with endogenous inputs, their share of influence can be non-zero. This phenomenon is better known as ``Shapley's joke'' \citep{ioocha22, Herin:2022} (or ``violation of the exclusion property'' \citep{Gromping:2007,ioocha22,cloioo23} or ``correlation distorsion'' \citep{verdinelli:2023}), and can be counter-productive for factor-fixing.
Therefore, \citet{Herin:2022} proposed to adapt the proportional values to the variance-based GSA paradigm, via the definition of the proportional marginal effects (PME) which are none other than the proportional effects of the dual. These indices also sum to $1$ and are always non-negative, but they bear an additional property: if, for a particular $A \in \pset{D}$, $X_A$ is the largest subset of exogenous inputs (in the sense that for any $B \in \pset{D}$ such that $A \subsetneq B$ contains endogenous inputs), then:
$$\textrm{PME}_i = 0, \forall i \in A.$$

Hence, if used in conjunction with the Shapley effects, the PME allow to not fall under the Shapley's joke, and hence strengthen the confidence in the resulting influence ranking. The PME differ from the Shapley effects, in the sense that the redistribution is not based on the same principles. From Eq.~(\ref{eq:shHarsanyi}), one can see that the Shapley effects provide an egalitarian redistribution of the Sobol' indices towards the involved subsets of inputs. The PME provide a proportional redistribution of the Sobol' indices: the more a player is worth in every coalition, the more wealth it receives \citep{Ortmann:2000, Feldman:2007, Gromping:2007}. This behavior is particularly obvious when dealing with two inputs, as pointed out in \citet{Gromping:2007, Herin:2022}.

\subsection{Estimation procedure, software and reproducibility}
\label{sec:estim}
As presented in \citet{Broto:2020}, a plug-in estimator of both Shapley effects and PME can be built using estimators of the total Sobol' indices $S^T_A$, for any $A \in \pset{D}$. Many estimation procedures have been proposed in the literature (e.g., based on Monte-Carlo sampling-based estimation \citep{Song:2016}). However, in our case study (see section \ref{study}), the input distributions are not known: only a single i.i.d input-output sample of PACS data is available. We therefore focus on 'given-data' estimators,  that do not require additional simulation.
More specifically, the method proposed by \citet{Broto:2020} is used, which allows for consistent estimators of the total Sobol' indices using a nearest-neighbor procedure (see also \citet{Idrissi:2021} for an overview of the method).

The following results have been gathered using the \texttt{sensitivity} R package, and in particular the \texttt{shapleysobol\_knn} and \texttt{pme\_knn} to estimate the Shapley effects and the PME respectively. This implementation allows categorical inputs, which are one-hot-encoded and normalized in order for the Euclidean distance to be sensible in the computation of the nearest-neighbors. To alleviate the computational burden of the nearest-neighbors procedure with many observations, they are computed using the approximation scheme proposed by \citet{Arya1998}.

Results and figures presented in the COVID-19 application can be reproduced with the codes available in a GitLab repository\footnote{See \url{https://gitlab.com/milidris/shapley-effects-and-pme-for-ct-scans}}.

\section{Comparison between PME and Shapley effects on a COVID-19 epidemic model}
\label{appli_covid}

The year 2020 raised important questions about the usefulness of epidemic modeling, especially in terms of producing relevant insights for public policy decision-makers, in light of the COVID-19 pandemic. \citet{Saltelli:2020} have used this example to emphasize the essential need for GSA in such models, which aim to predict the potential consequences of intervention policies.
Moreover, \citet{Lu:2023} highlighted that sensitivity analysis and uncertainty quantification were rarely performed in this context and that input interactions were often omitted. 
In the context of COVID-19 in Italy, \citet{Borgonovo:2020} proposed a first study to evaluate the sensitivity of key epidemiological model outputs, such as the number of individuals quarantined, recovered, or deceased due to COVID-19. Additionally, another GSA was conducted in the French context of the initial COVID-19 outbreak by \citet{daVeiga:2021livre}.
By using data coming
from this last analysis, \citet{daVeiga:2021livre} demonstrated how Shapley effects could be used to characterize the influence of uncertain input parameters on some outputs of this epidemic model. Using the same data, we highlight the differences between the PME and Shapley effects in this use-case example.

The deterministic compartmental model has been developed and described in \citet{daVeiga:2021livre} to represent the COVID-19 French epidemic (from March to May) by taking into account the asymptomatic individuals, the testing strategies, the hospitalized individuals and people admitted to Intensive Care Unit (ICU). Using several assumptions, it is based on a system of $10$ ordinary differential equations. Each equation models path of individuals between different compartments (corresponding to their infectious and illness states). These equations involve $20$ continuous input parameters $X$ assumed to be independent between each other and model the dynamic between the different compartments. In \citet{daVeiga:2021livre}, after a first screening step which allows for suppressing non-influential inputs, the model is calibrated on real data by using a Bayesian calibration technique. After the analysis of this step, the selected remaining inputs are:
$$X_{sel}=(p_a, N_a, N_s, N_{IH}, brn, t_0, M_u, N, Im_0)^T$$
Their joint probability distribution is approximated from a posterior sample derived from the Bayesian calibration process and the non-influential inputs are fixed to their nominal values. Table \ref{inputs_covid} presents the 9 selected input parameters. For the present study, we considered three output variables of interest: (i) the maximum value of U, with U the number of hospitalized patients in ICU; (ii) the total reported cases; (iii) the cumulative time where the maximum value of U is reached.

\begin{table}[ht!]
\centering
\begin{tabular}{c c } 
\hline
Input & Description \\ 
\hline
$p_a$ & Conditioned on being infected, the probability of having mild symptoms or non symptoms\\
$N_a$ & If asymptomatic, number of days until recovery\\
$N_s$ & If symptomatic, number of days until recovery without hospitalization\\
$N_{IH}$ & If severe symptomatic, number of days until hospitalization\\
$brn$ & Basic reproducing number\\
$t_0$ & Starting date of epidemics (in 2020)\\
$M_u$& Decaying rate for transmission (after social distancing and lockdown)\\
$N$ & Date of effect of social distanciation and lockdown\\
$Im_O$ & Number of infected undetected at the start of epidemic \\
\hline
\end{tabular}
\caption{Description of the uncertain input parameters of the COVID-19 epidemic model.}
\label{inputs_covid}
\end{table}

\begin{figure}[!ht]
\centering
\includegraphics[width=\linewidth]{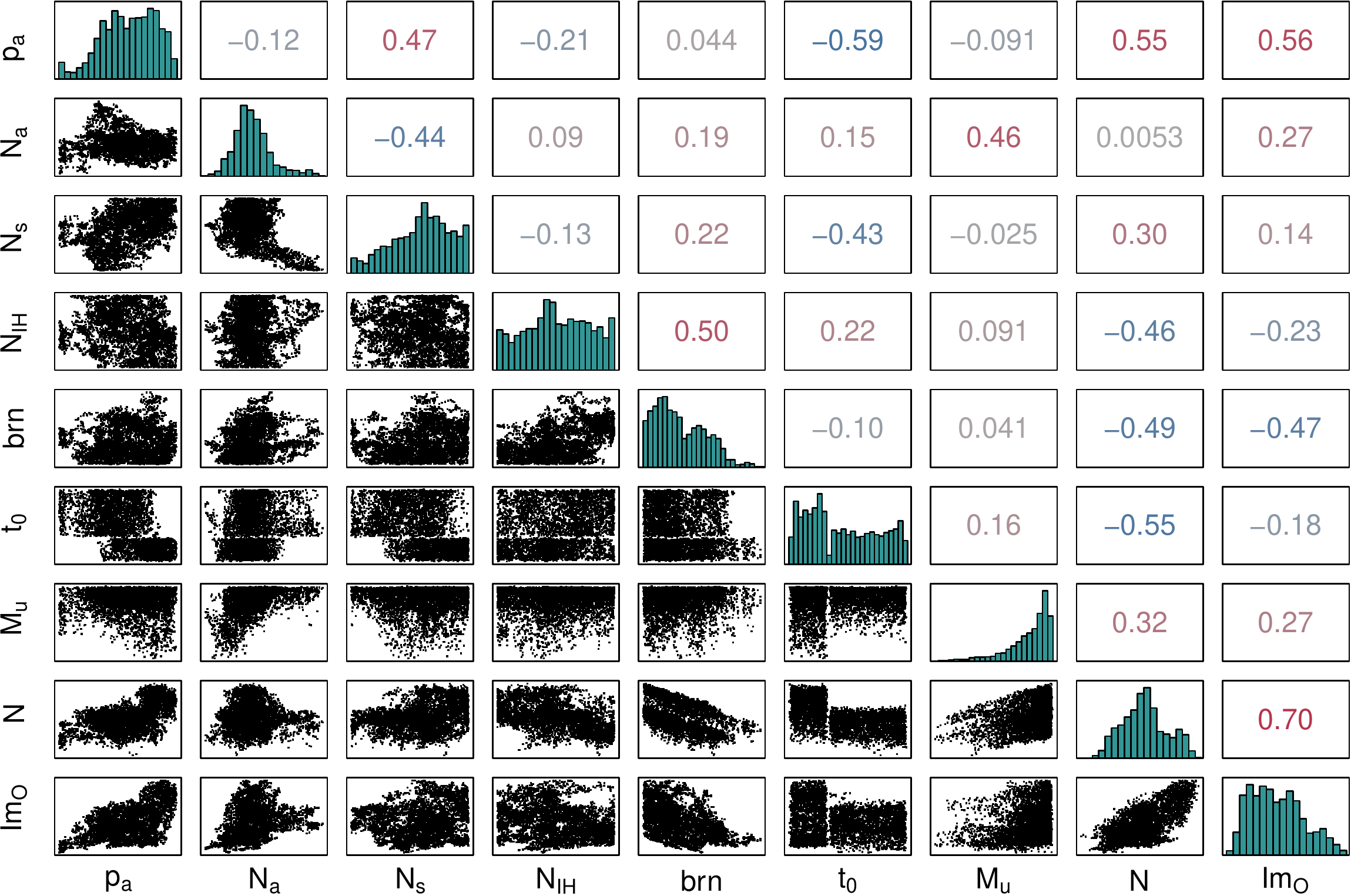}
\caption{Scatter-plots and estimated (Pearson's) linear correlation coefficients of the input parameters of the COVID-19 epidemic model, derived from a posterior sample.}
\label{correlcovid}
\end{figure}

Figure~\ref{correlcovid} displays estimated (Pearson's) the linear correlation coefficients between the input parameters selected after calibration. One can notice that some of these parameters can display strong correlations, and thus mutual independence is not satisfied. It comforts the choice of Shapley effects and PME in order to assess their influence. 

Shapley effects and PME have been computed from the methodology presented in Section~\ref{sec:estim}. The dataset consists in $5000$ rows, and thus, the approximated nearest-neighbor search algorithm has been used with a number of neighbors set fixed at $3$. The intervals around the estimated values are the empirical $2.5\%$ and $97.5\%$ quantiles, which have been computed by means of $100$ repetitions, where, for each repetition, the indices have been estimated on a random selection of $80\%$ of the initial data.

\begin{figure}[!ht]
\centering
\includegraphics[width=\linewidth]{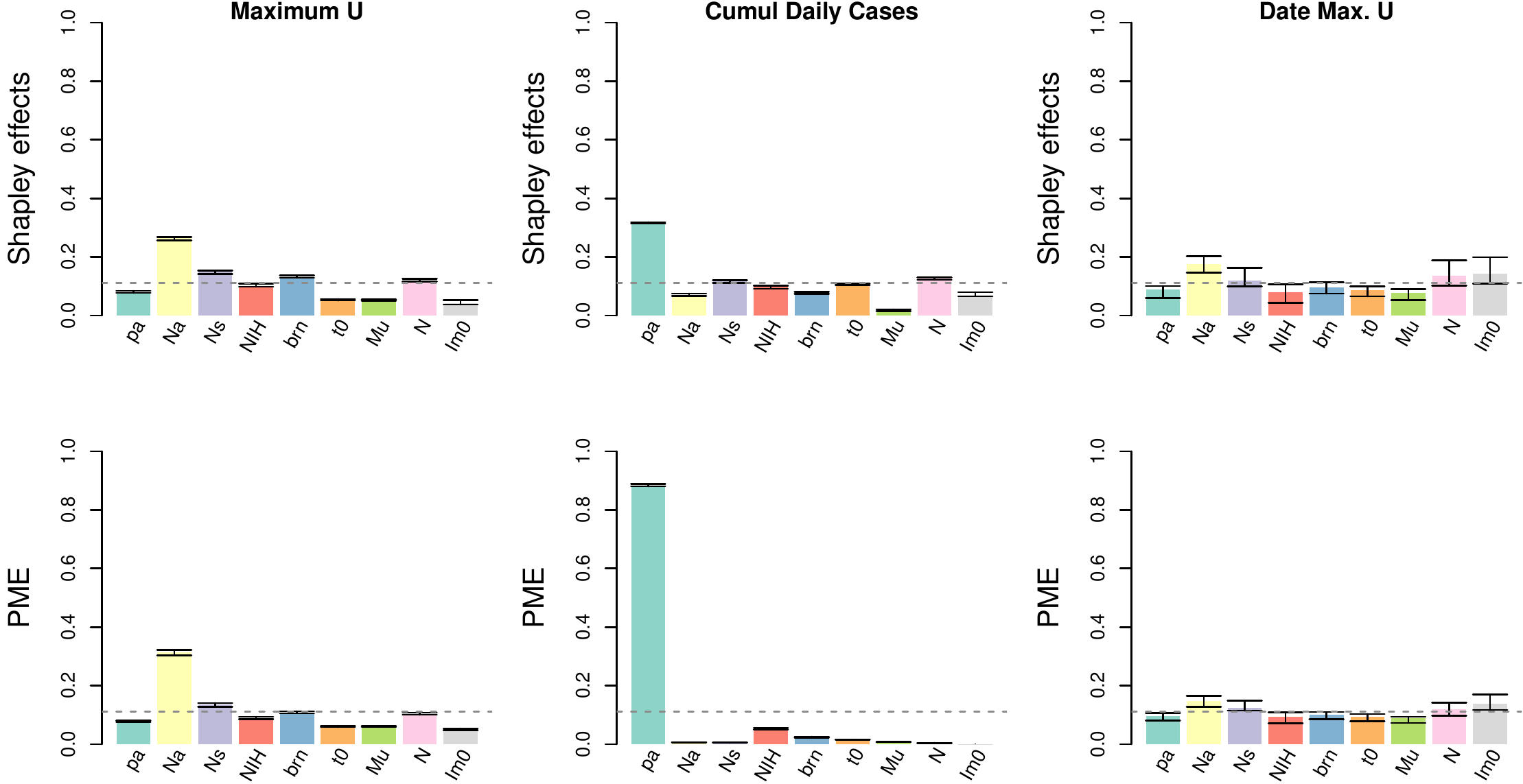}
\caption{Shapley effects and PME of 9 input parameters for three  outputs of the epidemic COVID-19 model: the maximum value of U, the total reported cases and the time where the maximum value of U is reached }
\label{shappmecovid}
\end{figure}

We observed that $N_a$ and $p_a$ were the most influential parameters of the COVID-19 epidemic model of interest. Shapley effects and PME estimated for the maximum value of U and the time where the maximum value of U is reached were similar (Table \ref{appli_covid}). However, according to Shapley effects, $p_a$ explained 30\% of the total reported cases variance whereas this parameter was granted more than 80\% of the variance share by PME, others parameters becoming non-influential (Figure \ref{shappmecovid}). $p_a$ was highly correlated with $t_0$, $N$ and $Im_O$ whereas $N_a$ was only moderately correlated with $N_s$ and $M_u$. 
For the total reported cases output, PME allowed for a better redistribution of $p_a$ interaction effects, proportionally to his marginal contribution and independently of his correlation structure. 

\begin{table}[!ht]
\centering
\begin{tabular}{c c c c c} 
\hline
Influence rank & \multicolumn{2}{c}{Shapley effects} & \multicolumn{2}{c}{PME}\\ 
\hline
& \multicolumn{4}{c}{Maximum U}\\
1 & $N_a$ &	26.68\% &	$N_a$ &	31.43\%\\
2 & $N_s$ & 14.98\% & $N_s$ & 13.50\% \\
3 & $brn$ & 13.17\% & $brn$ & 10.87\%\\
4 & $N$	& 12.01\% & $N$ & 10.36\%\\
5 & $N_{IH}$ & 10.06\% & $N_{IH}$ & 8.87\%\\
6 & $p_a$ &	8.02\% &	$p_a$ &	7.76\%\\
7 & $t_0$ &	5.45\% &	$t_0$ &	6.13\%\\
8 & $M_u$ &	5.41\% & $M_u$ &	6.10\%\\
9 & $Im_O$ & 4.21\% &	$Im_O$ & 4.97\%\\
\hline
& \multicolumn{4}{c}{Cumul Daily Cases}\\
1 & $P_a$ &	31.68\% &	$P_a$ &	88.35\%\\
2 & $N$ & 12.75\% & $N_{IH}$ & 4.91\% \\
3 & $N_s$ & 11.68\% & $brn$ & 2.61\%\\
4 & $t_0$	& 10.66\% & $t_0$ & 1.70\%\\
5 & $N_{IH}$ & 9.75\% & $M_u$ & 0.83\%\\
6 & $brn$ &	7.76\% &	$N_a$ &	0.61\%\\
7 & $N_a$ &	6.97\% &	$N_s$ &	0.53\%\\
8 & $Im_O$ & 6.76\% & $N$ &	0.46\%\\
9 & $M_u$ & 1.98\% &	$Im_O$ & 0.00005\%\\
\hline
& \multicolumn{4}{c}{Date Max. U}\\
1 & $N_a$ &	17.55\% &	$N_a$ &	14.78\%\\
2 & $Im_O$ & 14.32\% & $Im_O$ & 13.86\% \\
3 & $N$ & 13.62\% & $N_s$ & 12.35\%\\
4 & $N_s$	& 11.83\% & $N$ & 11.83\%\\
5 & $brs$ & 9.65\% & $brs$ & 10.05\%\\
6 & $p_a$ &	8.81\% &	$p_a$ &	9.52\%\\
7 & $t_0$ &	8.62\% &	$N_{IH}$ &	9.39\%\\
8 & $N_{IH}$ &	7.89\% & $t_0$ &	9.38\%\\
9 & $M_u$ & 7.71\% &	$M_u$ & 8.85\%\\
\hline
\end{tabular}
\caption{Influence hierarchies between the inputs of the COVID-19 epidemic model according to Shapley effects and PME.}
\label{hierarchy_covid}
\end{table}

\section{Application to CT scan dose estimation from the "restricted" NCICT model}
\label{appli_CT}

\begin{figure}[!ht]
\centering
\includegraphics[width=\linewidth]{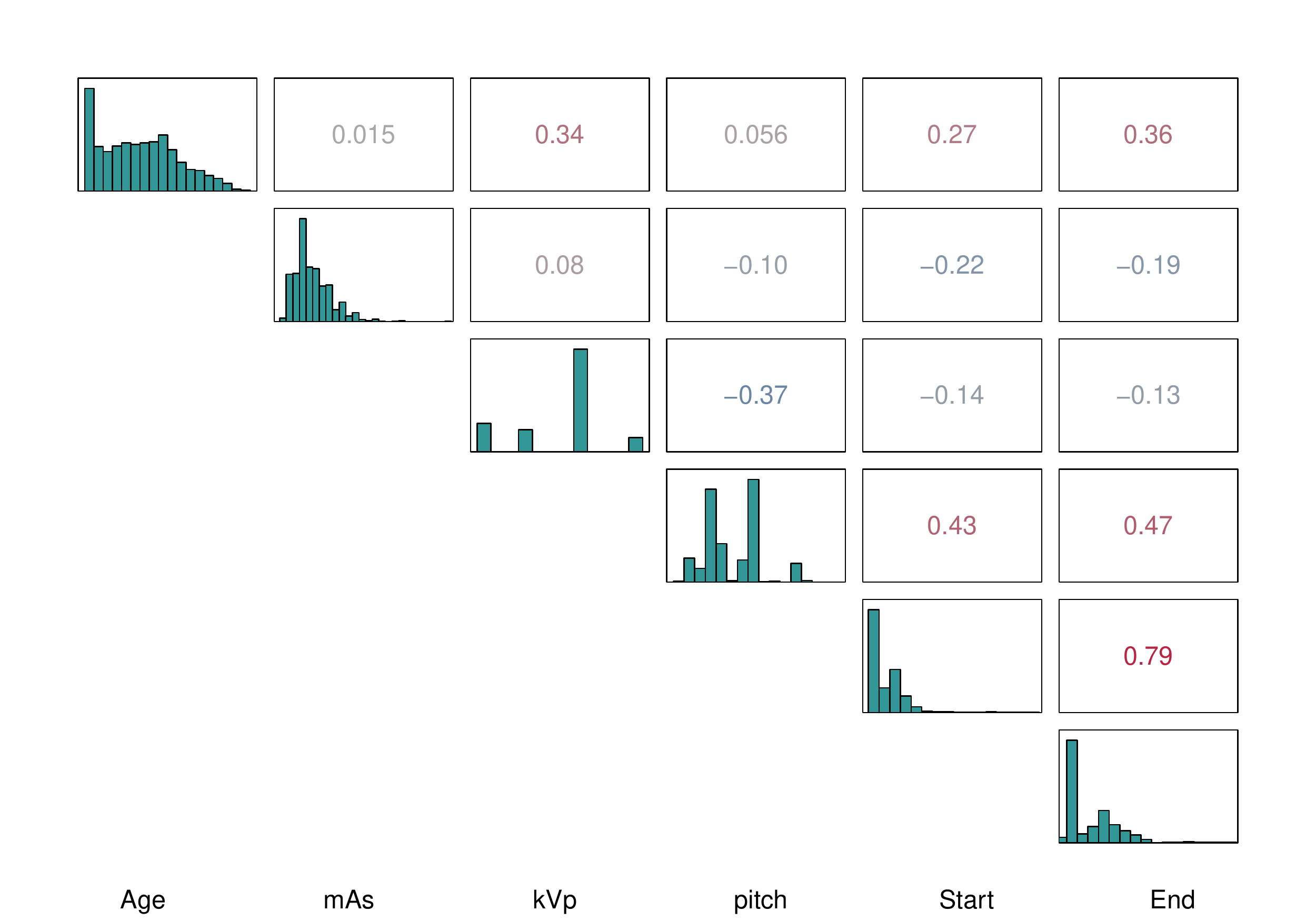}
\caption{Pairwise plots of the quantitative inputs of the restricted NCICT model including the histogram of the empirical distribution of each input (diagonal) and the estimated (Pearson's) linear correlation coefficients (upper part).}
\label{corrplot}
\end{figure}

Figure~\ref{corrplot} displays the estimated (Pearson's) linear correlation coefficients between the 8 input parameters of the "restricted" NCICT model used for CT scan dose estimation (see Table \ref{inputs}).
We observed relatively high correlation coefficients between scan start and scan end ($\hat{\rho}=0.79$), between pitch and scan start ($\hat{\rho}=0.43$), between pitch and scan end ($\hat{\rho}=0.47$) or between pitch and kVp ($\hat{\rho}=-0.37$). These four parameters, kVp, pitch, scan start and end are those with the higher values of variance inflation factor (VIF) (Table~\ref{VIF}) which is a metric of multicollinearity (see, e.g. \citet{cloioo23}). However, the two qualitative input parameters, gender and CT model, were not correlated with the quantitative variables as depicted by the correlation ratio in Table~\ref{correlRatio}. This correlation structure does not allow for interpretable Sobol' indices, which motivates the use of Shapley effects and PME. 

\begin{table}[!ht]
\centering
\begin{tabular}{ccccccccc} 
\hline
& Age & Gender & Model & mAs & kVp & pitch & Start & End\\
\hline
VIF & 1.46 & 1.01 & 1.43 & 1.14 & 1.47 & 1.69 & 2.80 & 3.25 \\
\hline
\end{tabular}
\caption{Variance inflation factor for the input parameters of the restricted NCICT model.}
\label{VIF}
\end{table}

\begin{table}[!ht]
\centering
\begin{tabular}{ccccccc} 
\hline
& Age & mAs & kVp & pitch & Start & End\\
\hline
Gender & 0.002 & 0.001 & 0.0002 & 0.002& 0.009& 0.005\\
Model & 0.02 & 0.07 & 0.03 & 0.05 & 0.05 & 0.07\\
\hline
\end{tabular}
\caption{Correlation ratio between the quantitative input parameters and the two qualitative variables "Gender" and CT "Model" of the restricted NCICT model.}
\label{correlRatio}
\end{table}

\begin{figure}[!ht]
\centering
\includegraphics[width=14cm]{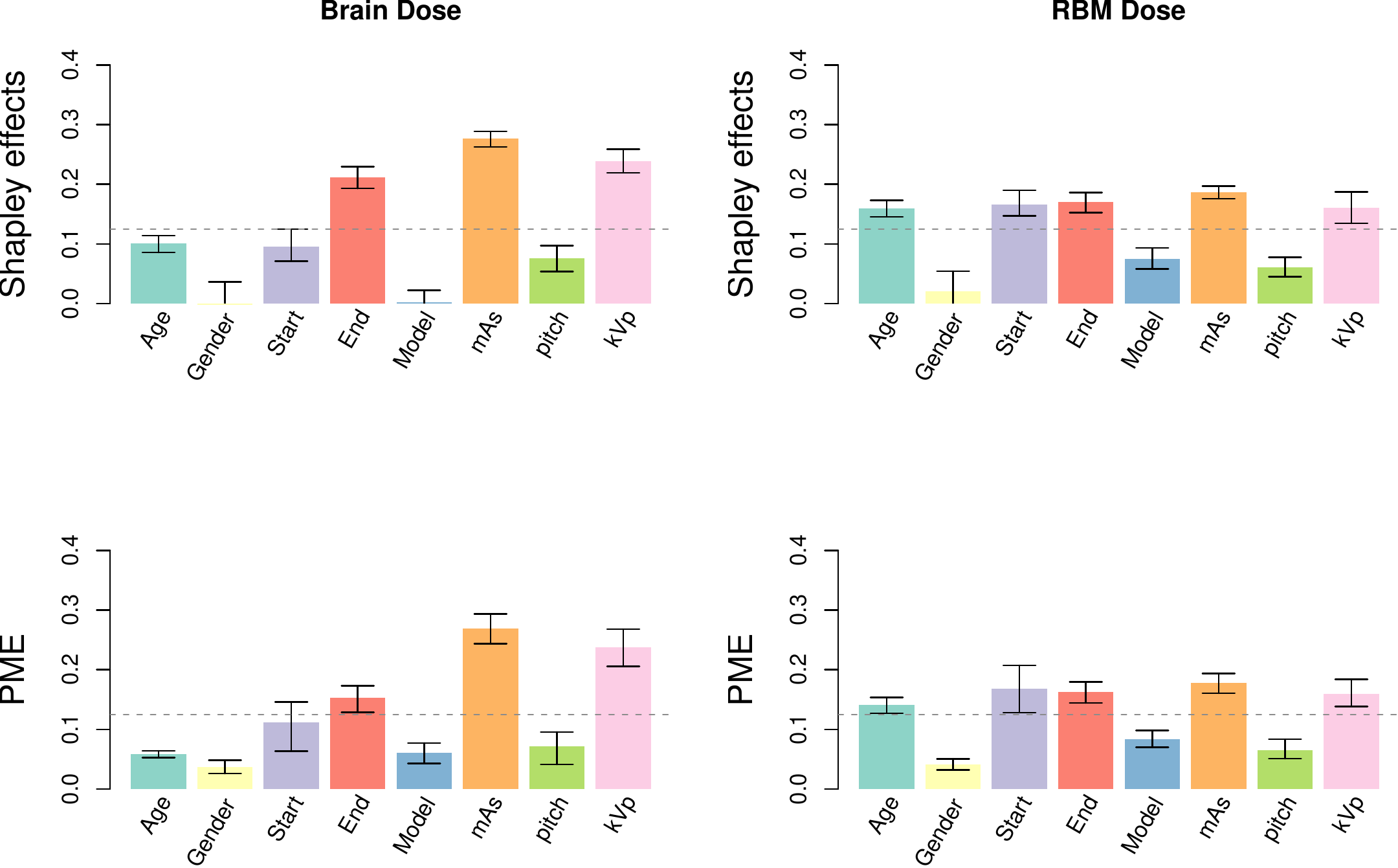}
\caption{Estimated Shapley effects (top) and PME (bottom) for brain (left) and RBM (right) dose estimates associated with head examinations.}
\label{head_fig}
\end{figure}

\begin{figure}[!ht]
\centering
\includegraphics[width=14cm]{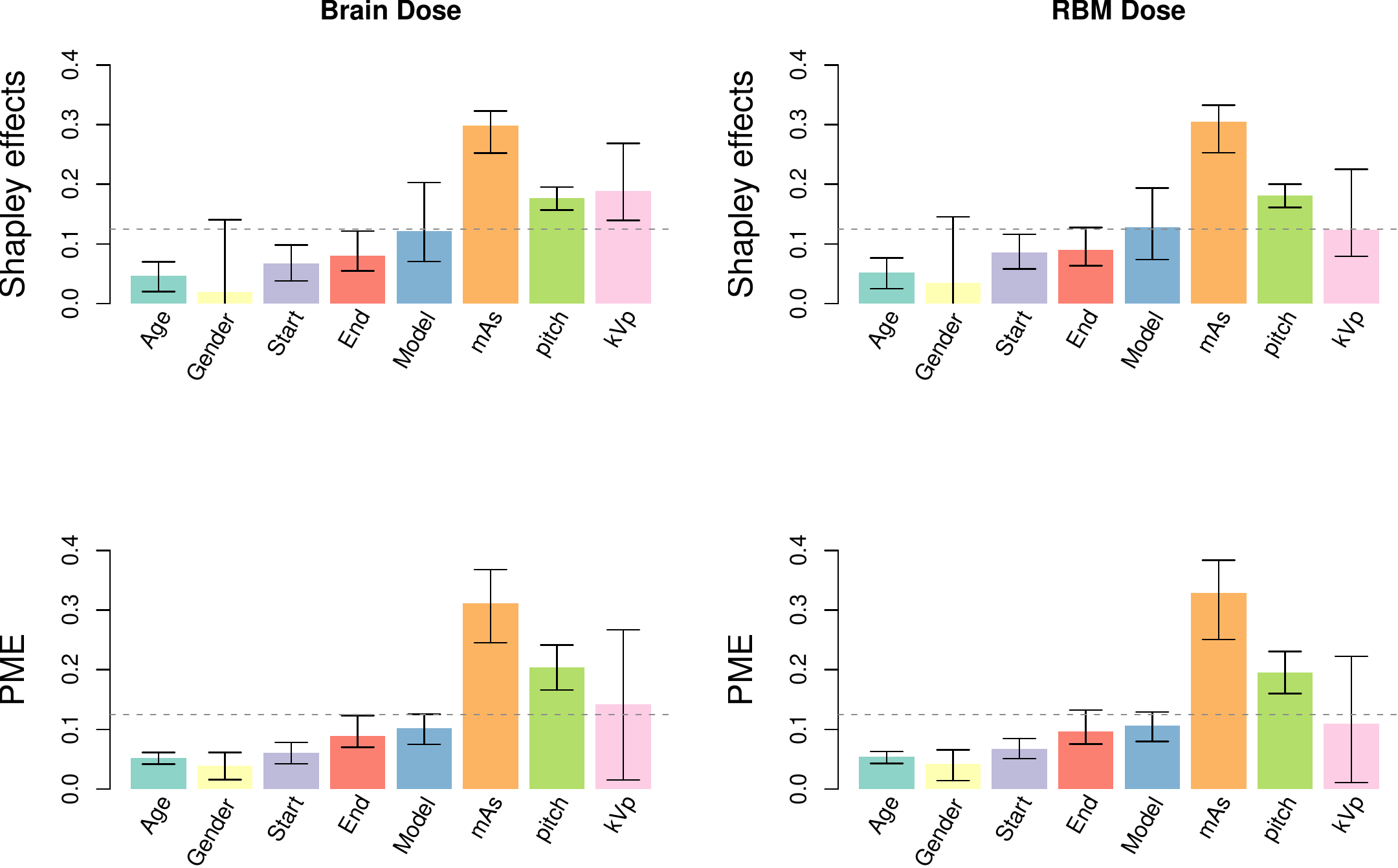}
\caption{Estimated Shapley effects (top) and PME (bottom) for brain (left) and RBM (right) doses associated with chest examinations.}
\label{chest_fig}
\end{figure}

\begin{figure}[!ht]
\centering
\includegraphics[width=14cm]{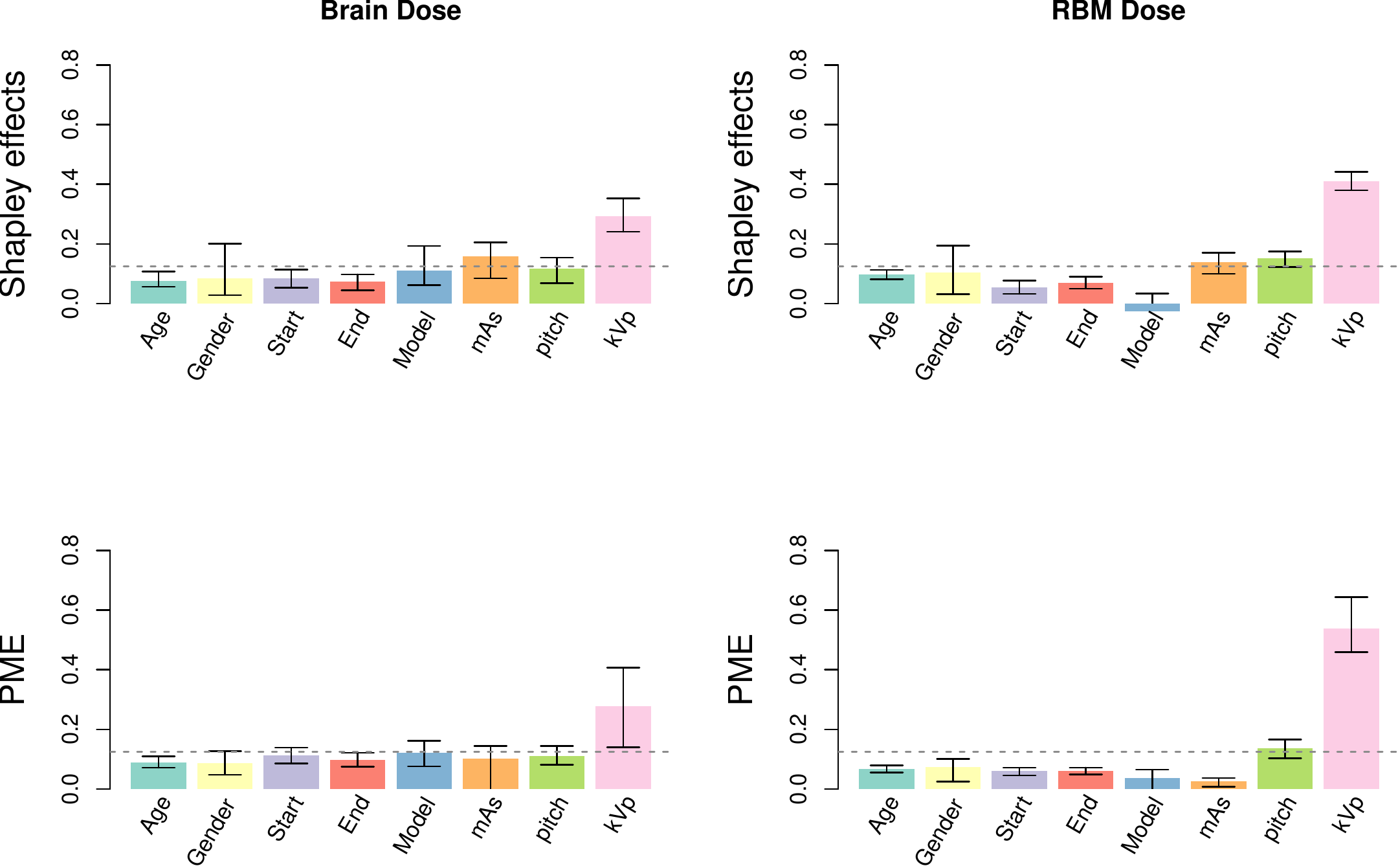}
\caption{Estimated Shapley effects (top) and PME (bottom) for brain (left) and RBM (right) doses associated with abdopelvis examinations.}
\label{abdopelv_fig}
\end{figure}

\begin{figure}[!ht]
\centering
\includegraphics[width=14cm]{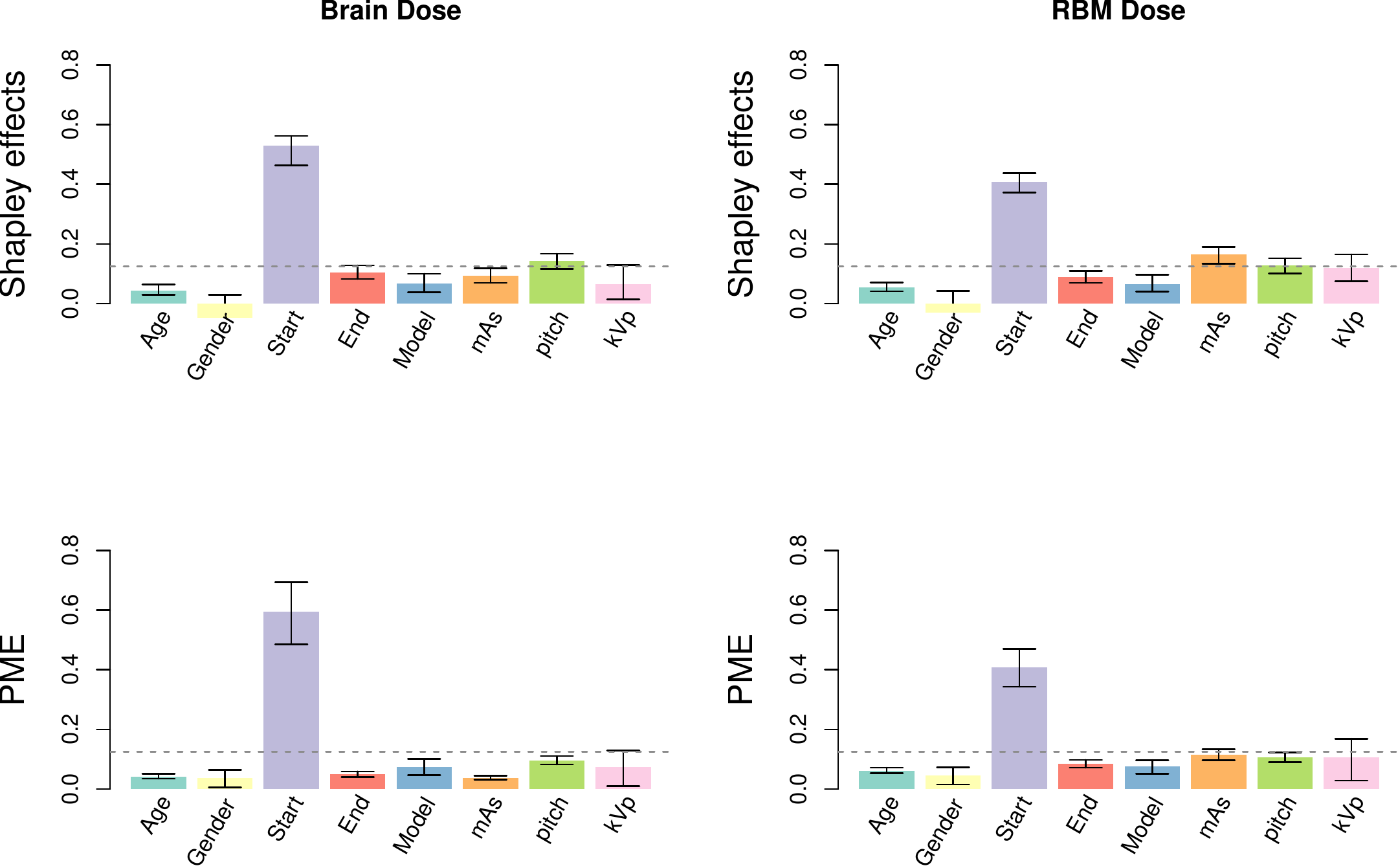}
\caption{Estimated Shapley effects (top) and PME (bottom) for brain (left) and RBM (right) doses associated with multiple areas examinations.}
\label{multiple_fig}
\end{figure}

Shapley effects and PME have been computed from the methodology presented in Section~\ref{sec:estim}. The dataset consists in 8848 rows and the approximated nearest-neighbor search algorithm has been used with an arbitrary number of neighbors set at N=100. The effects were forced to sum to one because of the deterministic nature of the model (i.e., there is not a noisy coefficient). The intervals around the estimated values are the empirical 5\% and 95\% percentiles, which have been computed by means of 200 repetitions, where, for each repetition, the indices have been estimated on a random selection of 90\% of the initial data, and subsequently corrected for the bias due to this sampling procedure. Figures \ref{head_fig}, \ref{chest_fig}, \ref{abdopelv_fig} and \ref{multiple_fig} display the Shapley effects and PME estimates for head (Number of examinations=4681), chest (N=2314), abdopelvis (N=823) and multiple areas (N=1025) CT scan examinations respectively. The dotted line represents the average influence of an input in the case of similar importance (i.e. 1/8 $\approx$ 12.5\%). Shapley and PME hierarchies are presented in Table~\ref{hierarchy}. The PME did not detect exogenous input since all allocations were strictly positive. It is reassuring since every input is effectively involved in the 
estimation of the organ dose. 

\textbf{Head examinations:} Figure \ref{head_fig} displays Shapley effects and PME for head examinations. On the one hand, the brain dose was mainly explained by mAs, kVp and scan end. Shapley effects and PME depicted similar values for mAs and kVp with around 27\% and 24\% of the brain dose variance respectively for these 2 parameters which are related to the quantity of delivered X-rays (Table~\ref{hierarchy}). However, scan end influence was lower with PME compared to Shapley effects, probably because of its correlation with other input parameters such as scan start ($\hat{\rho}=0.79$) and pitch ($\hat{\rho}=0.47$). On the other hand, for RBM dose, all input parameters were close to the average level, although five parameters seemed to be a bit more influential: mAs, kVp, scan start, scan end and age. It was not surprising that age intervened here since RBM distribution varies across the body depending on age, being less and less in the head as children grow up. 

\textbf{Chest examinations:} According to both estimated Shapley effects and PME displayed in Figure \ref{chest_fig}, mAs, pitch and kVp were the most influential input parameters of brain dose delivered during chest examinations. However, the PME allocated a lower share of the output variance to kVp than Shapley effects (Sh=19\%, PME=14\%) and a higher share to pitch (Sh=18\%, PME=20\%) (Table~\ref{hierarchy}). This could be explained by the high level of multicollinearity of these two input parameters with the other ones (VIF = 1.47 for kVp and 1.69 for pitch). Likewise, pitch is linearly correlated with scan start and end  ($\widehat{\rho}$ = 0.43 and 0.47). For RBM dose, mAs and pitch were the two influential parameters according to both sensitivity indices. In this case, the marginal contributions of these two parameters seem to be highly discriminated as the proportional redistribution of the interaction effects did not impact importance values compared to the Shapley effects.

\textbf{Abdopelvis examinations:} Regarding the brain dose, kVp appeared to be the only input parameter with a significant influence, that is to say greater than the average level of influence of one input (Figure \ref{abdopelv_fig}). However, its effect was uncertain, especially for PME. This uncertainty could be explained by the negligible brain dose delivered by abdopelvis CT examinations. Likewise, kVp was the most influential input parameter of RBM dose, explaining about half of the RBM dose variance (Table~\ref{hierarchy}). Furthermore, for brain dose, mAs was the second parameter in the Shapley hierarchy and the fifth in the PME hierarchy and for RBM dose, mAs was the third one in the Shapley hierarchy and the last one in the PME hierarchy. This parameter was not correlated with any other input but it interacts with other technical paramEters such as kVp and pitch to define the quantity of delivered X-rays. Thus PME allowed for a better redistribution of their interaction on dose estimates, proportionally to their marginal contributions (Table~\ref{hierarchy}).

\textbf{Multiple areas examinations:} Scan start was the most influential input parameter of doses estimates associated with multiple areas examinations (Figure \ref{multiple_fig}). It was granted half of the brain dose variance and 41\% of the RBM dose variance (Table~\ref{hierarchy}). This result could be explained by the heterogeneity of the considered examinations and scanned body regions included in this class. We observed some differences between Shapley effects and PME. For instance, according to Shapley effects, pitch explained 14\% of the brain dose variance and mAs explained 17\% of the RBM dose variance but these two parameters were allocated a lower influence by PME (Table~\ref{hierarchy}). 
Scan start was correlated with scan end and pitch. Compared to Shapley effects, PME was less influenced by inputs correlation and allocated a share of interaction effects to scan start proportionally to his marginal effect. This trend was also observed for the total reported cases of COVID-19 (Section \ref{appli_covid}). Finally, with Shapley effects, gender was granted a negative share of the doses variances which is due to the approximation error of the nearest-neighbors procedure. 

\begin{table}[!ht]
\centering
\begin{tabular}{c c c c c c c c c} 
\hline
Influence rank & \multicolumn{4}{c}{Brain dose} & \multicolumn{4}{c}{RBM dose}\\
& \multicolumn{2}{c}{Shapley effects} & \multicolumn{2}{c}{PME} & \multicolumn{2}{c}{Shapley effects} & \multicolumn{2}{c}{PME} \\ 
\hline
&\multicolumn{8}{c}{Head examinations}\\
1 & mAs & 27.6\% & mAs & 27.0\% & mAs & 18.7\% & mAs & 17.8\%\\
2 & kVp & 23.9\% & kVp & 23.7\% & End & 17.0\% & Start & 16.8\%\\
3 & End & 21.2\% & End & 15.4\% & Start & 16.6\% & End & 16.3\%\\
4 & Age & 10.1\% & Start & 11.2\% & kVp & 16.1\% & kVp & 16.0\%\\
5 & Start & 9.6\% & Pitch & 7.2\% & Age & 16.0\% & Age & 14.1\%\\
6 & Pitch & 7.5\% & Model & 6.1\% & Model & 7.5\% & Model & 8.3\%\\
7 & Model & 0.3\% & Age & 5.8\% & Pitch & 6.1\% & Pitch & 6.6\%\\
8 & Gender & -0.2\% & Gender & 3.7\% & Gender & 2.1\% & Gender & 4.2\%\\
\hline
&\multicolumn{8}{c}{Chest examinations}\\
1 & mAs & 29.9\% & mAs & 31,1\% & mAs & 30.6\% & mAs & 32.9\%\\
2 & kVp & 18.9\% & Pitch & 20,4\% & Pitch & 18.1\% & Pitch & 19.5\%\\
3 & Pitch & 17.7\% & kVp & 14,2\% & Model & 12.8\% & kVp & 11,0\%\\
4 & Model & 12.2\% & Model & 10,2\% & kVp & 12.3\% & Model & 10.6\%\\
5 & End & 8.0\% & End & 8.9\% & End & 9.0\% & End & 9.6\%\\
6 & Start & 6.7\% & Start & 6.1\% & Start & 8.5\% & Start & 6.8\%\\
7 & Age & 4.6\% & Age & 5.2\% & Age & 5.2\% & Age & 5.4\%\\
8 & Gender & 2.0\% & Gender & 3,9\% & Gender & 3.4\% & Gender & 4.2\%\\
\hline
&\multicolumn{8}{c}{Abdopelvis examinations}\\
1 & kVp & 29.4\% & kVp & 27,8\% & kVp & 41.0\% & kVp & 53,9\%\\
2 & mAs & 15.9\% & Model & 12,1\% & Pitch & 15.1\% & Pitch & 13.7\%\\
3 & Pitch & 11,6\% & Start & 11.3\% & mAs & 13.8\% & Gender & 7,3\%\\
4 & Model & 11.2\% & Pitch & 11,2\% & Gender & 10.3\% & Age & 6.8\%\\
5 & Start & 8.6\% & mAs & 10,1\% & Age & 9.7\% & End & 6,1\%\\
6 & Gender & 8.5\% & End & 9,8\% & End & 7,0\% & Start & 6,0\%\\
7 & Age & 7.5\% & Age & 9,0\% & Start & 5.5\% & Model & 3.7\%\\
8 & End & 7.3\% & Gender & 8.8\% & Model & -2.6\% & mAs & 2.6\%\\
\hline
&\multicolumn{8}{c}{Multiple areas examinations}\\
1 & Start & 53,1\% & Start & 59.6\% & Start & 40.8\% & Start & 40.8\%\\
2 & Pitch & 14.3\% & Pitch & 9.5\% & mAs & 16.5\% & mAs & 11.5\%\\
3 & End & 10.4\% & Model & 7.3\% & Pitch & 12.8\% & Pitch & 10.6\%\\
4 & mAs & 9.4\% & kVp & 7.3\% & kVp & 11,9\% & kVp & 10,6\%\\
5 & Model & 6.8\% & End & 4,9\% & End & 8.9\% & End & 8,3\%\\
6 & kVp & 6.5\% & Age & 4.1\% & Model & 6.5\% & Model & 7.6\%\\
7 & Age & 4.4\% & mAs & 3.8\% & Age & 5.5\% & Age & 6.1\%\\
8 & Gender & -4.8\% & Gender & 3.6\% & Gender & -3.0\% & Gender & 4.4\%\\
\hline
\end{tabular}
\caption{Influence hierarchies between the inputs of the restricted NCICT model according to Shapley effects and PME.}
\label{hierarchy}
\end{table}

\section{Discussion and conclusion}
\label{discu}

This paper provides a first comparison on a real medical application case, based on CT scan organ dose estimation, of the two recent variance-based GSA indices especially adapted to the dependent input parameters context: the Shapley effects \citep{Owen:2014} and the Proportional Marginal Effects (PME) \citep{Herin:2022}. Shapley effects were first defined to allocate a share of the output variance of the model to each input parameter, in the context of dependent inputs. Then, PME have been recently proposed to detect exogenous inputs. While Shapley allocation is egalitarian, the PME allocation tends to favor each input proportionally to its marginal contribution.

The aim of our case study in radiation dosimetry was to rank by influence the input parameters of the restricted NCICT model classically used to estimate the organ doses arising from CT scan exposure.
A GSA, based on the estimation of both Shapley effects and PME, allowed us to identify the most influential input parameters implied in brain and RBM doses estimation. Analyses were performed for four classes of examinations classified according to the scanned body region: head, chest, abdopelvis and multiple areas examinations. Among the eight uncertain input parameters implied in dose estimation, different influential parameters were highlighted depending on the studied scan type. For head examinations, mAs, kVp and the end of the scanned body contributed to about 70\% of the variance of the brain dose estimate while RBM dose was influenced by mAs, kVp, start, end and age. Brain and RBM doses delivered during chest examinations were mainly influenced by mAs, pitch and kVp. kVp was granted more than 30\% of the variance of the organ doses associated to abdopelvis examinations. Finally, the start of the scanned body was the only influential input parameter of brain and RBM doses related to multiple areas examinations. Thus, we identified two classes of influential parameters: (i) mAs, kVp and pitch, which represent the technical image acquisition parameters; (ii) the bounds of the scanned body region (scan start and end).

Additionally to the case study in radiation dosimetry that motivated this work, Shapley effects and PME have been applied to a use-case example (a COVID-19 epidemic model) in this paper. The goal was to better understand the behavior and usefulness of these two GSA indices.
The conclusion is that Shapley effects and PME are intrinsically different, but complementary. Shapley effects provide a tool for model exploration to identify the inputs that might have an impact on the output variance, even if it is due to correlation or interaction with other inputs. Initially, PME is rather focused on screening and factor fixing. In our case study in radiation dosimetry, all uncertain input parameters were endogenous and included in the NCICT equations. However, interestingly, we observed that the proportional redistribution property of the PME allowed for a clearer importance hierarchy: the PME are less sensitive to correlations and redistribute interaction effects proportionally to marginal effects.
Note also that this paper focused on the total contribution of each input but it does not provide information on the marginal contribution and the interactions between inputs. In a future work, the application of the Shapley-Owen indices \citep{rabbor19} may provide information about the interaction effects between inputs. Moreover, the present work focused on GSA relative to the variance of organ dose estimates. Inputs influence could also be studied from a different perspective and with other quantities of interest, such another statistical dispersion metrics or reliability-related quantities (e.g., a quantile), see \citet{daVeiga:2021livre} for an overview.

These two variance-based GSA indices, Shapley effects and PME, represent promising tools for forthcoming applications in epidemiology and, more generally, in operation research studies based on decision-support models. Indeed, mathematical models, which depend on many uncertain and dependent input parameters, are omnipresent in health studies, as in engineering, finance, management science, etc. \citep{bor17,razjak21}.
We have shown that, while Shapley effects allow to identify the most influential parameters, PME allow for a more pronounced hierarchy.
Thus, the comparison of these two allocations offer the possibility to understand marginal and correlation effects of the uncertain input parameters in a large variety of application domains of GSA.
Beyond GSA, these sensitivity indices are also promising in the context of interpretability of machine learning techniques \citep{leppal22}, as recently shown in \citet{ioocha22}.
The Shapley effects and PME are fully complementary in this context, depending if the aim is to perform an efficient screening (with PME) or to measure the effect of each considered potential input (of the machine learning model) fully accounting for their correlation induced effects (with Shapley effects).

\bibliographystyle{plainnat}
\bibliography{bib/references}
\end{document}